\documentclass[final,5p,authoryear,english,letter]{elsarticle}
\usepackage{geometry}
\usepackage[english]{babel}
\usepackage{graphicx}
\usepackage[authoryear]{natbib}
\usepackage{amsthm,amsmath}

\makeatletter
\def\ps@pprintTitle{%
   \let\@oddhead\@empty
   \let\@evenhead\@empty
   \let\@oddfoot\@empty
   \let\@evenfoot\@oddfoot
}
\makeatother

\begin{document}
\begin{frontmatter}

\title{Genetic Algorithms for Starshade Retargeting in Space-Based Telescopes}

\author[add1]{Ho Chit Siu}
\author[add2]{Victor Pankratius}

\address[add1]{Massachusetts Institute of Technology, Department of Aeronautics and Astronautics\\*[0.1cm] \textbf{hoseasiu@mit.edu}\\*[0.3cm]}
\address[add2]{Massachusetts Institute of Technology, Kavli Institute for Astrophysics and Space Research\\*[0.1cm]  \textbf{pankrat@mit.edu}\\*[0.4cm] {\normalsize July 23, 2019}}

\begin{abstract}
Future space-based telescopes will leverage starshades as components that can be independently positioned. Starshades will adjust the light coming in from exoplanet host stars and enhance the direct imaging of exoplanets and other phenomena. In this context, scheduling of space-based telescope observations is subject to a large number of dynamic constraints, including target observability, fuel, and target priorities. We present an application of genetic algorithm (GA) scheduling on this problem that not only takes physical constraints into account, but also considers direct human suggestions on schedules. By allowing direct suggestions on schedules, this type of heuristic can capture the scheduling preferences and expertise of stakeholders without the need to always formally codify such objectives. Additionally, this approach allows schedules to be constructed from existing ones when scenarios change; for example, this capability allows for optimization without the need to recompute schedules from scratch after changes such as new discoveries or new targets of opportunity. We developed a specific graph-traversal-based framework upon which to apply GA for telescope scheduling, and use it to demonstrate the convergence behavior of a particular implementation of GA. From this work, difficulties with regards to assigning values to observational targets are also noted, and recommendations are made for different scenarios.
\end{abstract}


\end{frontmatter}

\section{Introduction}
\subsection{Background and Motivation}
\begin{figure}
\begin{centering}
\includegraphics[width=0.5\textwidth]{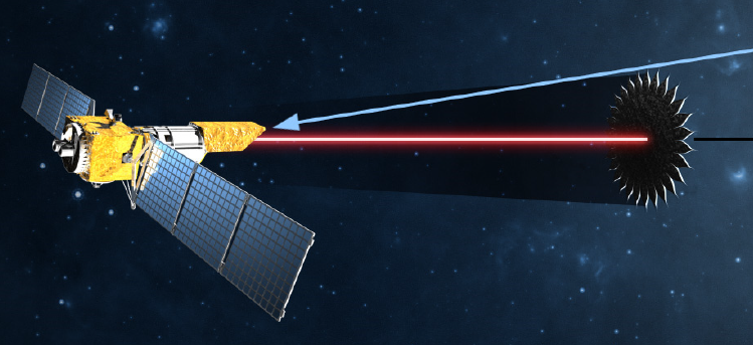}
\par\end{centering}
\caption{\label{fig:ExoS} Exo-S mission starshade telescope concept  \citep{exoSreport}.}
\end{figure}

The Exo-S Exoplanet Direct Imaging Mission Concept explored the possibility of using a space-based telescope along with a starshade to create the conditions necessary to directly image exoplanets \citep{exoSreport}. In this type of mission, a space-based telescope avoids the seeing effects of the Earth's atmosphere, while a starshade located in front of the telescope blocks the light of an exoplanet host star so that exoplanet(s) can be imaged. Separating the starshade from the telescope enables relatively small telescopes to be used to detect Earth-sized exoplanets due to the significant increase in the planet-star flux contrast, and the removal of the dependence of the system's inner working angle on the size of the telescope (Figure \ref{fig:ExoS}).

Since the starshade for such a mission is positioned 25,000 to 50,000 km from the telescope, a major difficulty associated with such a system lies in retargeting. The starshade must move large distances for each retargeting maneuver, meaning that each maneuver may take days to weeks to complete. Under such operating conditions, optimal scheduling becomes particularly important when trying to maximize science returns. Currently, the Exo-S mission divides potential targets into three tiers of priority, with targets of highest priority (systems with known giant planets and short characterization times) being scheduled first, followed by the two lower tiers (systems with observable habitable zones, and systems that are candidates for sub-Neptune or giant planet detection), which add targets in between observation of high-priority targets as allowable \citep{exoSreport}.

\begin{figure}
\begin{centering}
\includegraphics[width=0.5\textwidth]{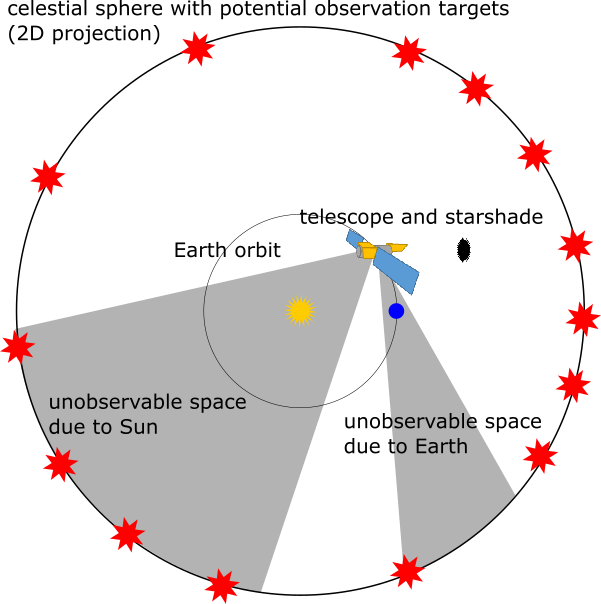}
\par\end{centering}
\caption{\label{fig:starshadeConstraintsExample} Simplified schematic of the time-varying constraints associated with scheduling a space-based Earth-leading starshade telescope. Not only do the unobservable regions change over time, the cost of retargeting is extremely expensive due to the distance that the starshade needs to traverse.}
\end{figure}

The starshade retargeting problem is difficult primarily due to the dynamic nature of its constraints. Figure \ref{fig:starshadeConstraintsExample} shows a 2D example of how the target selection for a mission in an Earth-leading orbit would be affected by the relative positions of the Sun and Earth. Since these constraints change over time, are complicated by the actual orbital dynamics involved in the mission, and involve potentially hundred or thousands of targets, we can quickly see that manual scheduling becomes extremely expensive and inefficient, and the problem even becomes intractable for most optimization frameworks. Added to these factors is the problem that observation campaigns may change based on newly collected data, which affects not only the amount of scientific interest in a given target, but also the cadence with which a target is best observed. Offering flexibility on the latter is common for observing campaigns using Earth-based observatories, but presents a much more significant challenge in the context of a space-based platform, particularly one with long retargeting times, and computer-aided discovery (\cite{pankratius2016}).

\subsection{Problem Statement}
In short, the problem presented by starshade scheduling may be stated as follows: given a set of observation targets of varying hypothetical scientific value, produce an observing schedule for a space-based telescope, subject to the environmental constraints of time-varying target visibility (e.g. due to the sun or other bodies being in the way) and mission constraints (e.g. fuel and time).

\section{Related Work}

\subsection{Scheduling}

The starshade retargeting problem is related in many ways to the well-known Traveling Salesman Problem, which involves finding the shortest path through a graph that visits every vertex given a set of edge distances. In the telescope context, the objective is to maximize the science gains of a mission, given a series of possible target stars to point to. Such problems are NP-hard, and in the worst case, have solving times that increase exponentially. In the starshade case, the action space grows roughly as $O(t^N)$ for $t$ discrete time steps and $N$ targets, even when we focus entirely on observation actions and ignore the additional complexities presented by the movements involved in retargeting. Due to their complexity, many probabilistic approaches to solving these types of problems have been proposed \citep{lin1965computer,lin1973effective,vcerny1985thermodynamical,dorigo1997ant}.

Alternatively, if retargeting is treated as a resource-constrained scheduling problem, we can see relationships with job-shop-type problems, where constraint satisfaction goes beyond simple task completion, and optimization is based on maximizing a reward function, rather like how we are trying to maximize science gains from the starshade mission. The scheduling of periodic and sporadic tasks in real-time systems \citep{chetto1989scheduling} is another field that this problem shares similarity with, since periodicity also exists in the starshade problem in the form of the orbit of the starshade around the Sun. However, in contrast to most of this kind of scheduling literature, we are not faced with periodic and aperiodic tasks that must be completed within a particular time, but rather, we see periodic constraints on the availability of tasks to be completed (i.e. variable availability of observational targets due to their angular distance from the Sun and other bright solar system objects), which is in ways the opposite of the kind of constraint that is more routinely explored in the literature. The starshade retargeting problem is thus an example of a subset of scheduling problems that is not as well-studied as the typical job shop or traveling salesman-type problem.

\subsection{Genetic Algorithms}

We propose a framework that poses the optimization as a graph traversal problem with arbitrary static and time-varying constraints to be solved by a genetic algorithm (GA). Genetic algorithms have been studied in the context of resource-constrained scheduling by \citet{wall1996genetic}, who developed a general approach to applying various types of constraints (such as precedence, resource availability, and dynamic variation) to a GA-based scheduler. Although he did not examine the effects of varying the parameters of the genetic algorithm, Wall found that GA-based approaches were flexible enough to perform well under arbitrary constraints. Since the flexibility of genetic algorithms is well-known, focus will be given to using a GA-based approach to the challenges faced by the Exo-S mission, and also to examining the effects of variations on the GA parameters. The flexibility of this approach, its implications for maximizing science returns, and the potential for human input on the scheduling problem will also be briefly discussed.

\begin{figure}
\begin{centering}
\includegraphics[width=0.5\textwidth]{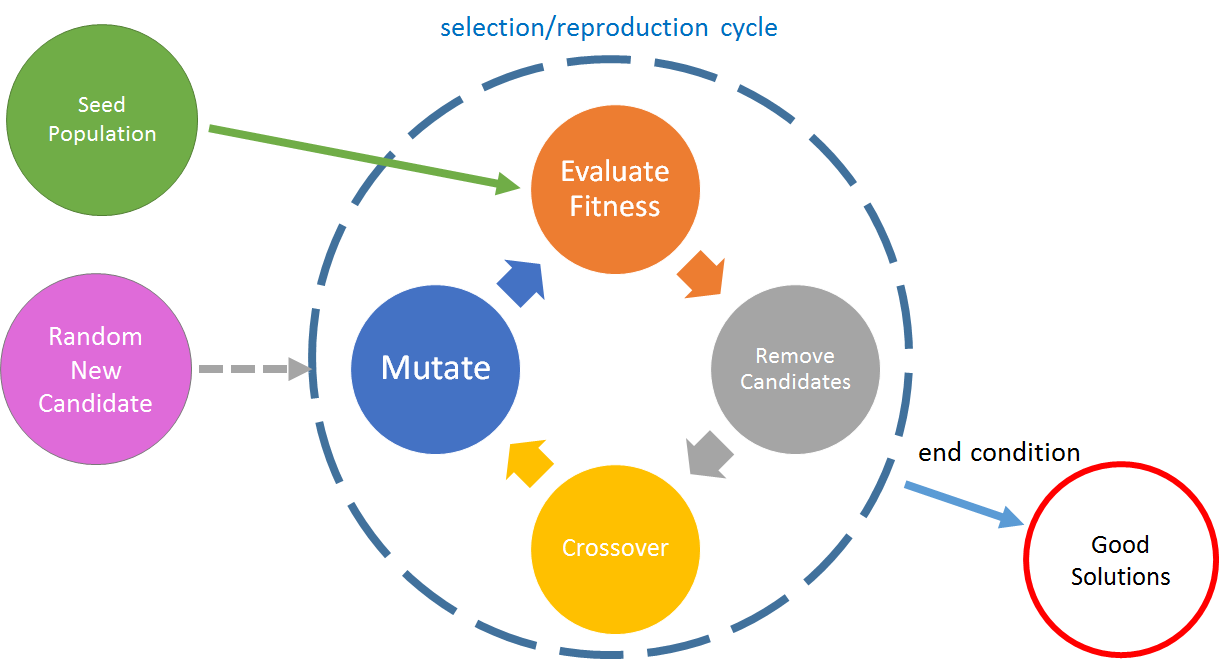}
\par\end{centering}
\caption{\label{fig:gaDiagram} Optimization process used in a genetic algorithm.}
\end{figure}

Genetic (or evolutionary) algorithms are a class of randomized algorithm that uses an optimization process similar to biological natural selection, as seen in Figure \ref{fig:gaDiagram} \citep{weise2009global}. Generally speaking, dimensions of the search space are represented as ``genes,'' and candidate solutions are ``individuals'' made up of multiple genes. An initial seed population of candidate solutions is randomly generated, and each candidate is evaluated using a fitness function. Low-fitness individuals are then removed from the population with removal function related to their fitness, signaling the end of a ``generation.'' The population is brought back up to its original size by ``mating'' surviving candidates --- that is, taking a combination of their parameters and producing a new candidate solution --- and applying random mutations to the resulting candidate. New candidates' fitness are then evaluated, and the cycle repeats until an end condition (usually some number of generations or a certain level of fitness) is met. In our implementation, we also allow for some probability of a random new candidate solution to take the place of one of the culled candidates at the start of a new generation in order to ensure genetic diversity.

\section{Problem Formulation}
The problem is formulated as a graph traversal, where a sequence of vertices in the graph represents the observing schedule. More formally, we specify a complete, undirected graph $G = \{V,E\}$ where,

\begin{itemize}
\item$V$ is a set of vertices, where each vertex $v$ represents an observation target with a ``science value'' $S(v,t_o)$, after the target has already been observed for a time $t_o$.

\item$E$ is a set of edges, where each edge $e=\{u,v\}$ is composed of an unordered pair of vertices $\{u,v\} \in V$, and has weights $F(e)$ and $T(e)$ representing the fuel and time costs associated with moving from $u$ to $v$. Since the graph is complete, $E$ contains all possible combinations of vertices.

\end{itemize}

The output of the problem is a \textit{walk} $W = {w_1, w_2, w_3 ... w_N}$, a finite ordered set representing the order and times in which targets are to be observed.

Several simplifications are made for this study. First, the problem is based  on the needs of the Exo-S Starshade Dedicated Mission concept, for which the telescope and starshade are dedicated to each other. This assumption couples the scheduling of the telescope and the starshade, and removes the necessity of considering science goals that might be achievable by the telescope alone.

Second, to eliminate the difficulty of assigning specific science value functions $S(v,t_o)$ to each target $v$, we assume an decreasing science value for each target per time step that it is observed in the form

\begin{equation}
S(v,t_o) = 0.5^{t_o} S_{0}(v),
\end{equation}

\noindent where the target starts out with a science value of $S_{0}(v)$ that is divided in half for each time step that it is observed (for a total observed time of $t_o$). Additionally, it is assumed that each target will be observed continuously until its science value reaches a predetermined minimum, after which it is not re-observed. This latter assumption dramatically reduces the search space necessary for an output, from $O(t^n)$ to $O(n!)$ (assuming no constraint on mission time $t$), as the plan is now determined by a \textit{path}, a particular type of walk where no vertex is repeated. Clearly, this form of scientific value would not really hold true for actual observation, and so other, more empirically-grounded science value functions would need to be created for different types of observations.

Third, a time constraint is not imposed on the overall mission. This means that the overall mission length is dictated solely by fuel constraints, and whether or not all targets have been observed. Time constraints are, however, still imposed for target observability and retargeting transit times, with the former being periodic constraints and the latter being dictated by the amount of movement necessary for a retargeting maneuver.

Fourth, the orbital dynamics of the problem are largely ignored. Our formulation essentially takes a fixed, telescope-centric reference frame for the celestial sphere, and ignores the blocking effects of all celestial bodies except for the Sun. Maneuvers are assumed to simply be direct arcs on the celestial sphere, and fuel costs are proportional to the length of the arc. The Sun's movements in Earth-centered RA and Dec are assumed to hold true for the telescope-centered reference as well, and any constraints on observability are treated as binary (completely observable or completely unobservable) based on the angular distance of the target from the Sun at any given point in time. Observations were deemed impossible if the target was within an arbitrary value of $60^\circ$ of the Sun (a value which will actually vary in reality depending on the type of observation being done). An observation start date of 1 January 2024 is assumed for the purposes of the solar constraint.

Finally, we assume that the telescope is always either observing or retargeting --- that is, the schedule does not allow for waiting, maintenance, or other idle activities. These activities could be added in when conducting a more detailed analysis, either to expand the search space (by waiting) or to impose additional constraints (by adding scheduled periods for downlink, software updates, or other tasks).

This formulation poses the problem as an optimization over a highly nonlinear, non-smooth function without derivatives. The lack of derivatives might normally point towards the use of stochastic methods such as gradient descent, simulated annealing \cite{aarts1988simulated}, or basin hopping \cite{wales1997global}, but even these methods assume smooth functions for optimization, making them inapplicable to this problem without significant modification. Genetic algorithms, on the other hand, are able to function with all of these constraints on the problem.

\section{Genetic Algorithm Specification} \label{sec:GAspec}

Since genetic algorithms are a class of optimization algorithms that follow a general framework, rather than a specific algorithm, there are a multitude of different ways to specify them, ranging from elements that exist in any particular instance in the entire class, such as the existence of candidate solutions, the population size, fitness functions, etc, to elements that may vary widely from instance to instance, such as different types of mutations, multiple populations, random restarts, chromosomal evolution, etc \cite{weise2009global}. Here, we take a relatively simple GA approach to produce an approximately-optimal schedule from the graph-based problem formulation, which is specified by the tuple $\{G,n,m,r\}$, where

\begin{itemize}
\item $G$ is the complete undirected graph specified in the problem formulation.

\item $p$ is the \textit{pool size}, representing the number of candidate paths in each generation.

\item $m$ is an array of size 3, containing the probability of a substitution mutation, an insertion mutation, and a deletion mutation for each crossover.

\item $r$ is the probability that each new path generated is a completely random new path as opposed to one created by crossing two existing paths.
\end{itemize}

We initialize the process by generating a seed population (set) $C$ of candidate schedules $c_i$ of size $p/2$. This half-population is generated at random by selecting targets and adding them to a schedule until no more could be added without fuel or time constraint violations. In the current restricted case, the amount of time spent at each target is fixed, because the telescope is set to observe each target until the amount of remaining science value for that target is below a certain minimum. Since the science value of targets can only decrease in this scenario, such targets are deemed completely observed and are not revisited. After the first-generation population is half-filled, the rest of the population is filled via a mating or random-introduction process.

For mating, pairs of randomly-selected schedules are selected to produce a ``child'' schedule via crossover and mutation. In crossover, the parent schedules are combined to form the basis of the child. A copy of each parent schedule first is turned from a directed graph to an undirected graph. Next, the pair of undirected graphs is composed into a single undirected graph by joining them at common nodes (removing the duplicates and joining the edges to the remaining node). If no common nodes exist, an edge is created to join one node on one parent graph to another node on the other parent. A random walk is then performed on the combined graph, starting at a random location and ending when a dead end is reached.

The directed graph created from this random walk is then mutated, with some independent probability of undergoing insertion, deletion, or substitution mutations based on the values in $m$. In an insertion mutation, a node not already in the graph is randomly inserted at an edge, and in a deletion mutation, a random existing node is removed, with the adjacent nodes being joined by a new edge. For a substitution mutation, a node is deleted, and then, if there exist one or more nodes that are not already in the original graph, one of these is selected at random to replace the deleted node. If the graph already contains all available nodes, two nodes are swapped. For each mating step process, we also have the possibility of a completely random schedule being introduced rather than a child schedule, which takes place with probability $r$.

Detection of a homogeneous population is done during the process of parent selection. Since parents are drawn randomly from the population, if two parent paths are found to be identical, they are not mated, and new parents are redrawn. If this occurs more than 100 times, a random path is introduced into the pool, regardless of the random path rate hyperparameter.

Once the pool is completely filled, the generation is complete, and the culling process begins. Here, fitness is determined by the function

\begin{equation}
max(S(c)) \text{ s.t. } F(c) < F_{max}, c \in C
\end{equation}

\noindent for maximizing the science value collected by candidate path $c$. If more than one candidate path has the same science value, then a second function

\begin{equation}
min(F(c))
\end{equation}

\noindent is used as a tiebreaker, with the candidate using less fuel being the fitter candidate. For the restricted case explored here, a probabilistic culling approach is taken, where each candidate $i$ has a modified fitness score

\begin{equation}
f'_i = (\frac{f_i-1}{n})^2,
\end{equation}

\noindent where $f_i$ is the fitness rank of candidate $i$ in the current generation. A random number $s_{min}$ is also generated for each candidate such that

\begin{equation}
f'_{min} \in [0,1]
\end{equation}

\noindent with a uniform probability distribution. The survival function is a comparison between $f'_i$ and $f'_{min}$ in the form,

\begin{equation}
survival(i) :=
\begin{cases}
1, & \text{if} \ f'_i \leq f'_{min} \\
0,  & \text{if} \ f'_i > f'_{min}
\end{cases}
\end{equation}

\noindent where 1 indicates survival, and 0 indicates removal from the population. This kind of culling is done rather than simply removing lower-ranked schedules in order to maintain genetic diversity.

The next generation is then started, beginning again with the creation of paths until the pool is once again full. This process is repeated until some end condition is met. In the case of convergence testing where the optimal path is known, the creation of a candidate that has the known optimal is treated as the end condition. When the algorithm is applied towards a set of targets where the optimal is unknown (i.e., all actual use cases), the end condition is more likely to be a set runtime, a set end generation, a set number of generations since the last schedule improvement, or something similar.

\section{Genetic Algorithm Testing}

Due to the large number of hyperparameters that are control various aspects of the algorithm (e.g. $p$, $m$, $r$) this genetic algorithm is not easily analyzed mathematically, and must be tested empirically. Rather than optimizing the hyperparameters, which was differ from case to case, we instead present a more general test of the number of candidate solutions that must be generated to obtain a solution that is within a certain percent margin of a known global optimum, which is defined to be the maximum science value within the given constraints. This kind of testing is similar to controlling the number of candidates and measuring the quality of the results generated, but we flip the independent and dependent variables in order to use a set of targets for which we know the global optimum. The hyperparameters were kept constant, and are shown in Table \ref{tab:hyperparameters}.

\begin{figure*}
\begin{centering}
\includegraphics[width=\textwidth]{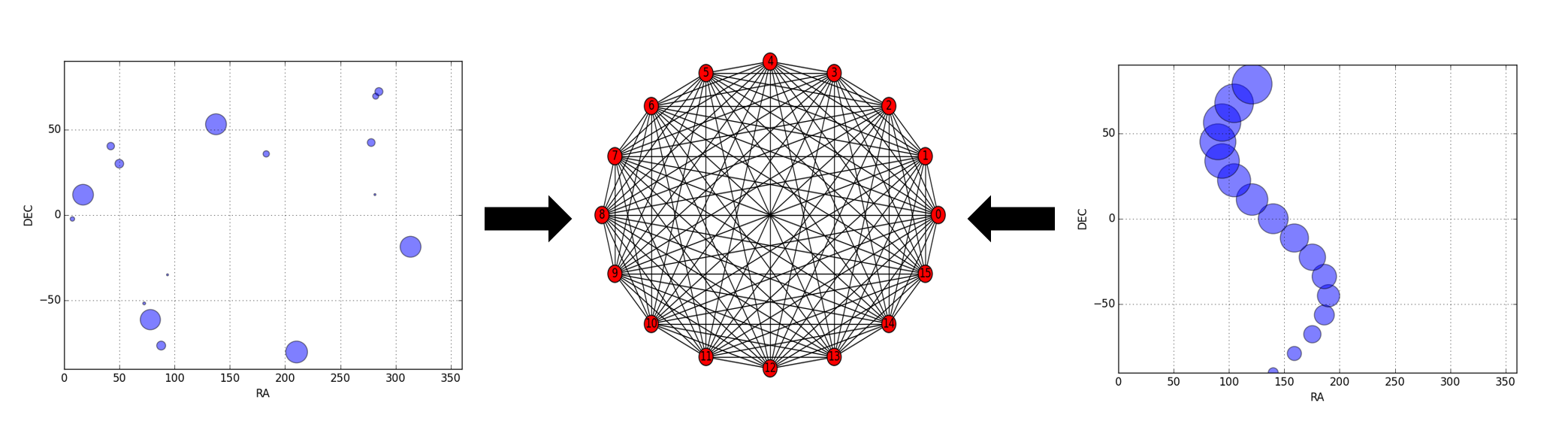}
\par\end{centering}
\caption{\label{fig:mapGraphExample} A random map of 16 targets (left), and the "curve" map used for testing (right). Size represents the science value of the target. Due to the graph framework being used for the GA, both maps are be converted to the topologically equivalent graph in the center, but with different vertex and edge weights.}
\end{figure*}

We constructed a scenario with 16 stellar targets arranged in a curve in RA-DEC space (Figure \ref{fig:mapGraphExample} right) without orbital constraints, but with the constraint that there is only enough fuel to reach 13 of the targets at most. This set of targets was designed specifically to avoid having to do an exhaustive search for a global optimum by being easy for a human to visually optimize: simply observe the targets in order from top to bottom, since they are already arranged spatially by science value. Despite the ease with which this case might be optimized for a human, it still shares the structure of a general case where target science values and locations are more random, as long as the number of targets is the same (Figure \ref{fig:mapGraphExample} left; note that RA wraps around), since to the genetic algorithm, both of these sets of targets are represented as a fully-connected graph (Figure \ref{fig:mapGraphExample} center). Both sets of targets are of equal difficulty to the genetic algorithm under this formulation, which is a practical benefit since actual target lists are unlikely to have locations that are visually well-structured.

\begin{table}[h]
\centering
\caption{GA hyperparameters.}
\label{tab:hyperparameters}
\begin{tabular}{l|l}
Hyperparameter 			& Value  \\ \hline
pool size (p)			& 64        \\
mutation rate (m)		& 0.15   \\
random path rate (r)	& 0.3	\\
iterations (i)			& 50
\end{tabular}
\end{table}

\section{Results} \label{sec:results}

\begin{figure}
\begin{centering}
\includegraphics[width=0.5\textwidth,height=8cm]{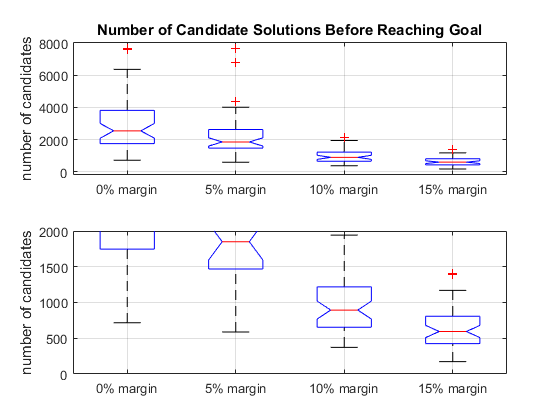}
\par\end{centering}
\caption{\label{fig:candidatesVsMargin} Number of candidate solutions evaluated before finding a schedule that is within a given percent margin of the optimal. Zero percent margin means that only the global optimum was accepted. The bottom plot is a zoomed-in version of the top plot.}
\end{figure}

\begin{figure}
\begin{centering}
\includegraphics[width=0.5\textwidth]{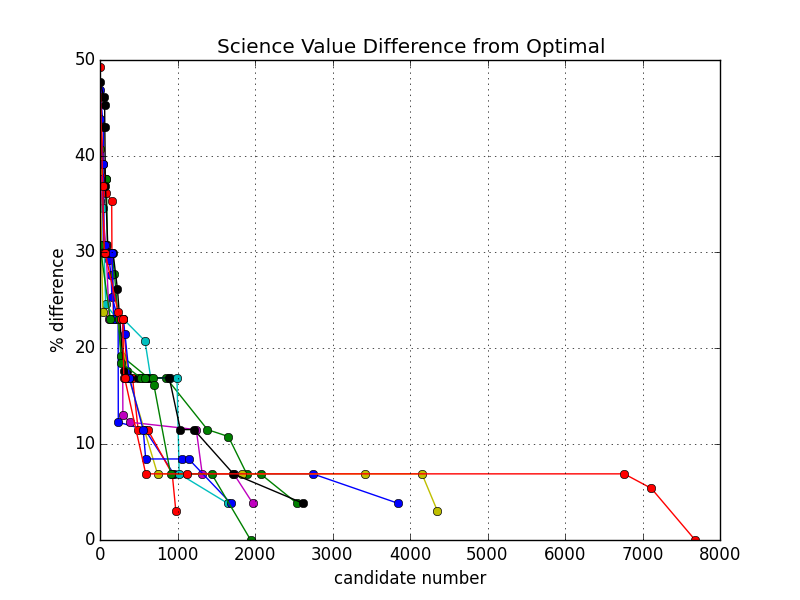}
\par\end{centering}
\caption{\label{fig:sciDelta} Percent difference in science value between the best solution at the time and the global optimum for 10 examples from the 50 runs in the 5\% allowable margin case. New points are only added to a line when the science value changes. Note the rapid convergence at the beginning, which slows down as more solutions are evolved and evaluated.}
\end{figure}

In Figure \ref{fig:candidatesVsMargin} we see that the number of candidate solutions evaluated decreases significantly as we relax the requirements of the optimization, meaning that true global optima would take much longer to find than approximate optima. Correspondingly, in Figure \ref{fig:sciDelta}, we see the change in the difference of proposed schedules' science values from the optimal schedule as a function of the number of candidates considered. Most of the change occurs at the beginning of the runs, within the first few generations, indicating that even if the algorithm were stopped before it reached the global minimum, it is likely that it would have already gotten fairly close to an optimal solution. All of the results show that evolving solutions using GA would take significantly fewer samples than a brute force method, which under this formulation has on the order of $16! = 10^{13}$ possibilities.

\section{\label{sec:discussion}Discussion}
\subsection{\label{subsec:challenges}Challenges of Using Genetic Algorithms}
We previously discussed a number of the benefits of using GA as a way to schedule space-based observation campaigns. In particular, we pointed out the applicability of this method to the highly nonlinear, non-smooth, possibly dynamic functions involved in the long repositioning time required in a starshade mission. However, the tests we conducted also demonstrate some of the challenges with using GA. Like most search methods, GA has a number of different hyperparameters that can affect its performance, and empirical testing is the main method for tuning these values. Particularly given its generational structure, there is also a tension between sampling enough of the search space by having large pools of candidates in each generation, or honing in on promising schedules more quickly be going through more generations, both of which add to runtime. More generally speaking, even optimality is only probabilistically guaranteed, though we showed that most of the gains of using GA are quickly realized before reaching optimal solutions.

\subsection{\label{subsec:Recommendations-for-Science}Recommendations for Science Value Assignment}

The framework presented here allows for a full observing schedule to be generated that satisfies constraints on fuel, time, and observing availability. These three constraints may be objectively measured, but the input for how valuable a target is for the overall science mission is something that must be determined by the science team beforehand, since it is a necessary but subjective input to the optimization problem that effectively constitutes the heuristic function used for the optimization. Indeed, this is a heuristic that might --- and very likely will --- change over the course of a mission as new discoveries are made. There are two components to creating this heuristic: giving a target a per-observation-time scientific value, and deciding how the value changes over time or over observations. The former is necessary to determine the priority of targets in the schedule, while the latter ensures that we do not simply observe a single high-value target for the entire duration
of the mission, and also incorporates the constraints of the kind of scientific return we can get from different kinds of data. 

The Exo-S report notes the conflicting goals of survey completeness for a particular class of planet, and survey diversity, given constraints on mission fuel and time. There is also the difficulty of scheduling observation time for stars with known planets, along with stars for which exoplanets are not confirmed. To address these concerns, there are several recommendations for how to assign science values to targets, and how to structure their change over time or observations.

First, we consider the simple case of comparing science values between targets for a particular snapshot in time. In the case of strict mission priorities (i.e. if confirmed exoplanet targets must always take precedent over unconfirmed exoplanet targets) science value for higher-tier targets should be assigned such that the minimum of these values is worth more than the sum of all targets with lower priority. The minimum ensures that any schedule that incorporates even a single higher-priority target will take precedence over schedules that do not. A softer version of this constraint might place a proportionally higher science value on higher-priority targets, but not follow the strict minimum defined here. This version of the constraint makes it possible (though unlikely) that a schedule that is able to observe a large collection of low-priority targets is found to be better than a schedule that observes the high-priority target.

Since there is no absolute measure of scientific value per observation of a target, it is only useful to consider how the values relate to each other. In this context, it is easier perhaps to think about science values not as a one-to-one comparison, but rather as a one-to-multiple comparison. For example, barring any strict prioritization of targets, in order for a target to have a science value three times that of another target, a single observation (for some definition of time for an observation) of that target must be worth three observations of different targets with the same value as the lower-valued target. This way of thinking about science value may may quantification easier than if one were to consider whether one target was ``worth'' three times as much as another.

As for how these values might change over time, consideration must be given to what kind of science is being done, and what kind of target is being observed. Depending on the level of time discretization, observing a target for a single unit of time may not be long enough for the observing objective. If there is a discrete minimum observation time necessary for science to be done with a target, then it needs to be applied either as a forced minimum observing period for the telescope, or the science value function must have a step increase so that a positive return is only obtained after the minimum observing time.

How the science values change on a per-observation basis must also be considered. In this case, the essential consideration is the value of adding an additional data point to the existing data on a target. For some targets for which parameters are poorly constrained (e.g. we have only a few observations of a potential transit), additional data may be highly valuable, and a drop-off in value may not occur for some time. On the other hand, for instances where the spacing of observations is important (again, using the example of a very long transit), then the change in value may be more complex, perhaps dropping immediately after an observation, but then increasing again later on. Clearly, this aspect of assigning science values can be a very involved process, and consider many more aspects of the science being done than the simple comparison-based assignments previously discussed.

The assignment of science values to targets and the assignment of variation functions is likely to be one of the most difficult parts of implementing a GA-based approach to retargeting. Its subjective nature makes it potentially a much more contentious issue for different members of the science team on any starshade mission, in contrast to the optimization of GA parameters, which at least can be shown to have some approximate numerical optimum.

On one hand, making these assignments is not all that different than the observing scheduling that is done currently by humans, who must make judgments on the relative values of different observing proposals. On the other hand, the additional algorithmic layer added to the process, and the removal (to some degree) of direct human assignment of the observing schedule would likely make it more difficult to convince end users to adopt this approach as compared to manual or more deterministic scheduling. The recommendations presented here are just a small set of ways to give the algorithm a set of starting conditions that more closely matches the desires of science teams, but much like the way GAs generally only give approximately optimal solutions, satisfaction of competing science goals is likewise approximate, even given the best match of target science values and variations to the actual desires of scientists.

\section{Future Work}
Additional work on this method of mission scheduling can take several different forms. It is worthwhile to consider the applicability of the method described here to an actual set of targets rather than the constructed example shown here for demonstration. The Exo-S target list provides a good opportunity to have a set of targets with known locations, and some sense of scientific importance attached to them, though the use of that particular list is limited by export control restrictions. Other example lists or scenarios could also be constructed with specific emphasis on different tiers of scientific value, time dynamics, effect of new discoveries, multiple starshades, or other specific factors. Ultimately, much of the future work on this type of scheduling will be dependent on the specific nature of missions and mission proposals as they arise.

At the algorithmic level, different implementations of GA could also be explored, including previously-mentioned attributes such as random restarts, multiple populations, chromosomal evolutions, and the like \citep{weise2009global}, though like the cases presented here, these formulations must also be tested empirically.

\section{Conclusion and Outlook}
Genetic algorithms are a promising approach for the problem of scheduling space-based telescopes, and in particular, of scheduling telescope using starshades, which require long transit times between targets. Our graph-based problem formulation uses a number of simplifications such as a single set of GA hyperparameters. Despite its simplicity, we demonstrate the practical utility of this kind of graph formulation and show the general trends of how our algorithm approaches an optimal solution. Further, we discuss the challenges of creating a value system for observation targets and derive initial recommendations for how these details might be addressed for different types of targets. Further work on this approach will need to be highly tailored to the specific details of the mission to which it will be applied.

\section*{Acknowledgements}

\noindent We acknowledge support from NASA AISTNNX15AG84G and NSF ACI1442997.

\vspace{0.5cm}

\section*{References}
\bibliographystyle{plainnat}

\end{document}